\documentstyle[prd,aps,preprint,epsf,axodraw,psfig]{revtex}
\newcommand{\gsim}{ \mathop{}_{\textstyle \sim}^{\textstyle >} }
\newcommand{\lsim}{ \mathop{}_{\textstyle \sim}^{\textstyle <} }

\newcommand{\EV}{~\mbox{eV}}
\newcommand{\kEV}{~\mbox{keV}}

\newcommand{\GEV}{~\mbox{GeV}}
\newcommand{\TEV}{~\mbox{TeV}}

\begin{document}
\tighten
\draft

 \baselineskip 0.6cm
\renewcommand{\thefootnote}{\fnsymbol{footnote}}
\setcounter{footnote}{1}
 
\title{\hfill{\normalsize\vbox{\hbox{UT-966}
}}\\
Leptogenesis via $LH_{u}$ flat direction\\
with a gauged $U(1)_{B-L}$}
 \vskip 1.2cm
\author{Masaaki~Fujii, K.~Hamaguchi}
\address{Department of Physics, University of Tokyo, Tokyo 113-0033,
Japan}
\author{T. Yanagida}
\address{Department of Physics, University of Tokyo, Tokyo 113-0033,
Japan \\ and \\
Research Center for the Early Universe, University of
Tokyo, Tokyo, 113-0033, Japan}

\vskip 2cm
%\date{\today}
%
\maketitle

\vskip 2cm
\begin{abstract}
  
We study the supersymmetric leptogenesis via $LH_{u}$ flat direction
with a gauged $U(1)_{B-L}$ symmetry.  We find that the resultant baryon
asymmetry is enhanced compared with the case without a gauged
$U(1)_{B-L}$ symmetry.  The baryon asymmetry is proportional to the
reheating temperature of inflation, but it is independent of the
gravitino mass. If high reheating temperatures of inflation $T_{R}\sim
10^{10}\GEV$ are available, the mass of the lightest neutrino, $m_{\nu 1}\sim
10^{-4}\EV$, is small enough to explain the baryon asymmetry in the present
universe. Furthermore, the gravitino mass independence of the produced
baryon asymmetry allows us to explain the present baryon asymmetry even
in low energy SUSY breaking scenarios such as gauge-mediation models.
Our leptogenesis scenario is also very special in the sense that it is
completely free from the serious Q-ball formation problem.
\end{abstract}
\newpage
%\pacs{PACS numbers: ..., ..., ...}
%12.10.-g Unified field theories and models \\
%12.60.Fr Extensions of electroweak Higgs sector \\
%12.60.Jv Supersymmetric models (see also 04.65 Supergravity)\\
%11.25.Mj Compactification and four-dimensional models \\
%
\renewcommand{\thefootnote}{\arabic{footnote}}
\setcounter{footnote}{0}

%%%%%%%%%%%%%%%%%%%%%%%%%%%%%%%%%%%%%%%%%%%%
%%%%%%%%%%%%%%%%%%%%%%%%%%%%%%%%%%%%%%%%%%%%
%%%%%%%%%%%%%%%%%%%%%%%%%%%%%%%%%%%%%%%%%%%%
\section{Introduction} 
\label{sec:introduction}
%%%%%%%%%%%%%%%%%%%%%%%%%%%%%%%%%%%%%%%%%%%%
%%%%%%%%%%%%%%%%%%%%%%%%%%%%%%%%%%%%%%%%%%%%
Our universe is composed of baryon, but not anti-baryon.  Accounting for
this baryon asymmetry in the present universe is one of the most
fundamental and challenging problems in particle physics as well as in
cosmology.  Various mechanisms have been proposed so far to solve this
problem. In particular, there has been a growing interest in
leptogenesis~\cite{LG-org}, since there are now strong evidences of
neutrino oscillations, equivalently the existence of tiny but nonzero
neutrino masses. It is, then, very natural to consider lepton-number
violating interactions in nature, which is a crucial ingredient for
leptogenesis to work.  Especially, once we accept the supersymmetry
(SUSY) and the existence of tiny Majorana neutrino masses, there
automatically exist in the minimal supersymmetric standard model (MSSM)
most of the necessary ingredients for the SUSY leptogenesis
via $L H_{u}$ flat direction, 
which was originally suggested by Murayama and
one of the authors (T.Y.)~\cite{MY}\footnote{See also later
studies~\cite{DRT,MM}.}  based on the idea of the Affleck-Dine 
(AD) mechanism~\cite{AD}.
   
Recently, we performed a detailed analysis of this MY leptogenesis via
$L H_u$ flat direction in the context of gravity-mediated SUSY breaking
scenarios including all of the relevant thermal effects~\cite{AFHY,FHY}.
We found a very interesting aspect of this leptogenesis, that is,
``reheating-temperature independence of cosmological baryon
asymmetry''~\cite{FHY}.  The resultant baryon asymmetry is almost
determined by the mass of the lightest neutrino (and the gravitino
mass), and hence we can determine the lightest neutrino mass from the
baryon asymmetry in the present universe, quite independently of the
reheating temperature of inflation.  This results in a crucial
prediction on the rate of the neutrinoless double-beta
$(0\nu\beta\beta)$ decay~\cite{FHY}, which may be testable in future
$0\nu\beta\beta$ decay experiments such as GENIUS~\cite{GENIUS}.

In this article, we study the effects of gauging the $U(1)_{B-L}$
symmetry on the leptogenesis via the $LH_{u}$ flat direction. ($B$ and
$L$ are baryon and lepton number, respectively.) As well known, the
$U(1)_{B-L}$ symmetry is the unique global symmetry which can be gauged
consistently with the MSSM.  If we accept a gauged $U(1)_{B-L}$
symmetry, the existence of three families of the right-handed neutrinos
are automatically required from the anomaly cancellation conditions.
Furthermore, if we assume that this $U(1)_{B-L}$ symmetry is
spontaneously broken at high energy scale, these right-handed neutrinos
acquire large Majorana masses of the order of the $B-L$
breaking scale, which naturally explains the tiny neutrino masses
suggested from the recent neutrino oscillation experiments via the
so-called ``seesaw mechanism''~\cite{seesaw}.

Below the $B-L$ breaking scale, the scalar potential along the $LH_{u}$
direction is almost the same as in the case without a gauged
$U(1)_{B-L}$ symmetry. However, above the $B-L$ breaking scale, the
scalar potential is lifted by the effect of the $U(1)_{B-L}$ D-term in a
certain region of parameter space, which provides us a new
mechanism~\cite{ADwB-L} to stop the $\phi$ field.  ($\phi$ is the
superfield parameterizing the $LH_{u}$ ``effectively flat'' direction.)
We show that if the $\phi$ field is stopped by the D-term potential, the baryon
asymmetry is {\it enhanced} compared with the ``global'' case without a
gauged $U(1)_{B-L}$ symmetry. This is because the available A-term,
which provides a phase rotational motion of the $\phi$ field generating
the lepton asymmetry, is much bigger than that in the global case. On
the other hand, if the $\phi$ field is stopped by the F-term potential 
coming from
the operator responsible for generating the small neutrino mass, the
obtained baryon asymmetry becomes the same as that in the global
case~\cite{AFHY,FHY}. (In the following discussion, we will call the
former case ``D-term stopping case'', while the latter one ``F-term
stopping case.'')

Surprisingly enough, as we will see in the following discussion, if the
amplitude of the $\phi$ field is fixed by the D-term potential 
during inflation, the baryon asymmetry linearly depends on the reheating
temperatures of inflation, but it is completely independent of the
gravitino mass. This is totally an opposite situation to the 
global case, in which the final baryon
asymmetry is almost independent of the reheating temperatures of
inflation and linearly depends on the gravitino mass.

We find in the D-term stopping case that if high reheating temperatures,
$T_{R}\simeq 10^{10}\GEV$, are available, the lightest neutrino of mass,
$m_{\nu1}\simeq 10^{-4}\EV$, can account for the present baryon
asymmetry. Furthermore, the ``gravitino-mass independence of the
cosmological baryon asymmetry'' provides us a great advantage in
gauge-mediated SUSY breaking scenarios~\cite{GMSB}.  Gauge-mediated SUSY
breaking is widely considered as one of the most interesting mediation
mechanisms of SUSY breaking, since it can naturally explain the
suppression of the flavor changing neutral currents (FCNC). However, in
those scenarios, the reheating temperature of inflation is strongly
restricted from above in order to avoid the overproduction of
gravitinos~\cite{g-prob-GMSB}. This makes most of the baryo/leptogenesis
scenarios impossible, especially when the gravitino is lighter than
${\cal{O}(\mbox{MeV})}$, since in this case the reheating temperature
$T_{R}$ is restricted as $T_{R}\lsim 10^{5}\GEV$.\footnote{ The
electroweak baryogenesis seems very difficult to take place in
gauge-mediated SUSY breaking models because it is hard to obtain the
required light stop, $m_{\tilde{t}}<m_{t}$. } It was considered that the AD
baryogenesis~\cite{AD} (which is the baryogenesis based on the AD
mechanism) is the unique candidate to produce an enough baryon asymmetry
in gauge-mediated SUSY breaking scenarios. However, from a recent
detailed analysis, it becomes clear that the AD baryogenesis is not
successful in gauge-mediated SUSY breaking scenarios because of a
serious Q-ball formation~\cite{Q-prob-GMSB}. In the present work, we
will show that the MY leptogenesis
can naturally explain, in the D-term stopping case,  the present baryon
asymmetry even in gauge-mediated SUSY breaking models with the gravitino
mass smaller than ${\cal{O}}(100\kEV)$, if the mass of the lightest
neutrino is
$m_{\nu 1}\lsim 10^{-10}\EV$. We should stress that such a light
gravitino may  be directly detected in future collider
experiments~\cite{MakiOrito}.  We claim that our scenario is the first
minimal model which explains the present baryon asymmetry in the
gauge mediation model with  a small gravitino mass
$m_{3/2}={\cal{O}}(100\kEV)$.

%%%%%%%%%%%%%%%%%%%%%%%%%%%%%%%%%%%%%%%%%%%%%%%%%%%%%%%%%%%%
%%%%%%%%%%%%%%%%%%%%%%%%%%%%%%%%%%%%%%%%%%%%%%%%%%%%%%%%%%%%
\section{The model}
\label{sec:model}
%%%%%%%%%%%%%%%%%%%%%%%%%%%%%%%%%%%%%%%%%%%%%%%%%%%%%%%%%%%%%
%%%%%%%%%%%%%%%%%%%%%%%%%%%%%%%%%%%%%%%%%%%%%%%%%%%%%%%%%%%%%
First, let us discuss the potential for the relevant fields.  To
demonstrate our point, we use a simple
superpotential as follows:
\begin{equation}
W=\mu H_{u}H_{d}+hNLH_{u}+\frac{1}{2}\lambda SNN+\eta X\left(S\bar{S}-v^{2}\right)
\;.
\label{s1-1}
\end{equation}
Here, $h,\;\lambda,\;\eta$ are coupling constants and we assume
$\eta={\cal{O}}(1)$.  $X$, $S$ and $\bar{S}$ are singlets under the MSSM
gauge group and $S,\;\bar{S}$ carry $-2,\;+2$ of $B-L$ charges,
respectively.  $N,\;L,\;H_{u}$ and  $H_{d}$ are a right-handed Majorana
neutrino, an $SU(2)_{L}$-doublet lepton and Higgs fields which couple
to up and down type quarks, respectively.  $v$ is the breaking scale of
the $U(1)_{B-L}$ symmetry.\footnote{Here and hereafter, we take the
couplings $h$, $\lambda$, $\eta$ and $\mu$, $v$ to be real. This can be
done by field redefinitions in Eq.(\ref{s1-1}).}  As we will see later,
the ``effectively flat'' direction relevant for the present baryon
asymmetry is the flattest flat direction, and hence we will consider
only one family which corresponds to the lightest left-handed neutrino.
We adopt the following linear combination of $L$ and $H_{u}$ as the flat
direction field $\phi$~\cite{MY} :
\begin{equation}
L=\frac{1}{\sqrt{2}}\left(
\begin{array}{cc}
\phi\\
0\end{array}
\right),\qquad
H_{u}=\frac{1}{\sqrt{2}}\left(
\begin{array}{cc}
0\\\phi
\end{array}
\right)\;.
\label{s1-2}
\end{equation}

The K\"ahler potential must have non-minimal couplings of the $\phi$
field to the inflaton, otherwise the $\phi$ field gets a large positive
mass term of the order of the Hubble parameter (which we will call a
Hubble mass term) and it is driven exponentially towards the origin
during inflation, and hence the leptogenesis via $LH_{u}$ flat direction
cannot take place~\cite{DRT}. Therefore, we assume that there are
general non-minimal couplings of the $\phi$ field to the inflaton in the
K\"ahler potential,
\begin{equation}
\delta K\ni \left(\frac{c_{\phi}}{M_{*}}I\phi^{\dag}\phi+{\rm{h.c.}}
\right)+\frac{b_{\phi}}{M_{*}^{2}}\phi^{\dag}{\phi}I^{\dag}I+\ldots
\;.
\label{s1-3}
\end{equation}
Here $|c_{\phi}|,\;b_{\phi}={\cal{O}}(1)$ are coupling constants, and
$M_{*}\simeq 2.4\times 10^{18}\GEV$ is the reduced Planck scale.  $I$ is
the inflaton superfield. {}From these non-minimal couplings to the
inflaton, the $\phi$ field
 gets the following SUSY breaking effects
during inflation and at the inflaton-oscillation dominated epoch:
\begin{equation}
\delta V\ni -\sqrt{3}H (c_{\phi}\phi W_{\phi}+{\rm{h.c.}})+
3(1-b_{\phi}^{\prime})H^{2}|\phi|^{2}+\ldots
\;,
\label{s1-4}
\end{equation}
where $H$ is the Hubble parameter, $b_{\phi}^{\prime}\equiv b_{\phi} -
|c_{\phi}|^{2}$, $W_{\phi}\equiv\partial W/\partial \phi$, and we use
the fact that $|F_{I}|^{2}\simeq3 H^{2}M_{*}^{2}$ in these
regimes.\footnote{We redefined the phase of $c_{\phi}$.} In the
following discussion, we assume $3(1-b_{\phi}^{\prime})\simeq -1$ for
simplicity.  We also assume that there are non-minimal couplings of
other fields to the inflaton in the K\"ahler potential.  Then, the full
scalar potential relevant to the $\phi$ field is given by\footnote{Here,
we omit the potential coming from thermal effects, which will be
discussed in Section~\ref{Sec-BS} and \ref{Sec-5}.}
\begin{eqnarray}
V=&&\frac{1}{2}g^{2}\left(-2 |S|^{2}+2|\bar{S}|^{2}+|N|^{2}-|L|^{2}\right)^{2}
\nonumber\\
&&+|\mu H_{u}|^{2}+|h L H_{u}+ \lambda S N|^{2}
+|\mu H_{d}+h N L|^{2}+|h N H_{u}|^{2}\nonumber\\
&&+\left|\frac{1}{2}\lambda N^{2}+\eta X \bar{S}\right|^{2}
+|\eta(S \bar{S}-v^{2})|^{2}+|\eta X S|^{2} \nonumber\\
&&+3 H^{2}\left(\sum_{Y}(1-b_{Y}^{\prime})|Y|^{2}\right)
-\sqrt{3}H \left(\sum_{Y}c_{Y} Y W_{Y}+\rm{h.c.}\right)+V_{SB}
\;,
\label{s1-5}
\end{eqnarray}
where $b_{Y}^{\prime}\equiv b_{Y}-|c_{Y}|^{2}$, $c_{Y}$ are non-minimal
couplings of $Y(=L,H_{u},H_{d},X,S,\bar{S},N)$ to the inflaton in the
K\"ahler potential as coupling constants in Eq.~(\ref{s1-3}), and
$g={\cal{O}}(1)$ is the gauge coupling constant of the
$U(1)_{B-L}$. $V_{SB}$ represents soft SUSY breaking terms in the
present true vacuum.  Here, all fields denote the scalar components of
the corresponding superfields.

The scalar potential in Eq.~(\ref{s1-5}) is so complicated that it seems
very difficult to solve the dynamics of the relevant
fields. Fortunately, however, we only need to know the shape in
the neighborhood of the bottom of this potential, if the curvature
around this bottom is as large as the Hubble parameter during inflation. 
This is because all the scalar fields, which have masses as large as the
Hubble parameter, settle down at the bottom of the potential during
inflation and trace this potential minimum throughout the history of the
universe.  This allows us to eliminate many terms in the potential by
using the field relations required for minimizing the potential.
Therefore, let us first find out the conditions to minimize the
potential in Eq.~(\ref{s1-5}).  Though this is also a very hard task
because of the complexity of the potential, one can find out the minimum
of the potential at least when a condition $|\phi|,\;H\lsim v$ is
satisfied.  In fact, we find  the approximate minimum of F-terms for
$|\phi|, H\lsim v$,~\footnote{Here, we
assume a positive Hubble-order mass term for the $H_{d}$ field. This is
necessary in order to avoid contamination of $H_{d}$ to the $LH_{u}$
flat direction. We have checked both analytically and numerically that
the contamination of the $H_{d}$ field becomes nonnegligible only after
the $H\lsim \mu$, and hence the following discussion is not affected
much by the $H_{d}$ contamination.}
\begin{eqnarray}
\bar{S}\simeq \frac{v^{2}}{S},\quad X\simeq 0,
\quad N\simeq -\frac{h LH_{u}}{\lambda S},\quad H_{d}\simeq 0
\;.
\label{s1-6}
\end{eqnarray}
The curvatures around  the first three minimums are of the order of the
$B-L$ breaking scale $v$, and that around the last one is of the order of the
Hubble parameter.  As explained above, we simplify the
potential by using the relations in Eq~(\ref{s1-6}).  Then, the
potential in Eq.~(\ref{s1-5}) is  reduced to
\begin{eqnarray}
V\simeq &&\frac{1}{2}g^{2}\left(
-2|S|^{2}+2\frac{v^{4}}{|S|^{2}}+
\frac{h^{2}}{4 \lambda^{2}}\frac{|\phi|^{4}}{|S|^{2}}
-\frac{1}{2}|\phi|^{2}\right)^{2}
+\frac{h^{4}}{4\lambda^{2}}\frac{|\phi|^{6}}{|S|^{2}}+
\frac{h^{4}}{64\lambda^{2}}\frac{|\phi|^{8}}{|S|^{4}}
\nonumber\\
\nonumber\\
&&-H^{2}|\phi|^{2}
+3(1-b_{S}^{\prime})H^{2}|S|^{2}+3(1-b_{\bar{S}}^{\prime})H^{2}\frac{v^{4}}{|S|^{2}}
+3(1-b_{N}^{\prime})H^{2}\frac{h^{2}}{4\lambda^{2}}\frac{|\phi|^{4}}{|S|^{2}}
\nonumber\\
\nonumber\\
&&+\frac{\sqrt{3}}{2}H \left(c_{\phi}^{\prime}\frac{h^{2}}{\lambda}\frac{\phi^{4}}{S}
+\rm{h.c.}\right)+\frac{1}{2}\mu^{2}|\phi|^{2}+V_{SB}
\;.
\label{s1-7}
\end{eqnarray}
Here, $c_{\phi}^{\prime}\equiv c_{\phi}-1/4\; c_{S}$.  One may wonder
why the ``flat'' direction field $\phi$ can develop a large expectation value
in spite of the presence of the $U(1)_{B-L}$ D-term.  This is because
the $S$ field shifts and absorbs the D-term potential. Here, we have
\begin{equation}
|S|^{2}\simeq -\frac{1}{8}|\phi|^{2}
+\sqrt{v^{4}+\left(\frac{1}{64}+\frac{h^{2}}{8\lambda^{2}}\right)|\phi|^{4}}
\;.
\label{s1-8}
\end{equation}
Then, we get the following potential:
\begin{eqnarray}
V\simeq &&V_{SB}+\frac{1}{2}\mu^{2}|\phi|^{2}+
\frac{\sqrt{3}}{2}H\left(c_{\phi}^{\prime}\frac{h^{2}}{\lambda}
\frac{\phi^{4}}{S}+\rm{h.c.}\right)+\frac{h^{4}}{4\lambda^{2}}
\frac{|\phi|^{6}}{|S|^{2}}+{\cal{O}}\left(\frac{|\phi|^{8}}{|S|^{4}}\right)
\nonumber\\
\nonumber\\
&&-H^{2}|\phi|^{2}
+3(1-b_{S}^{\prime})H^{2}|S|^{2}+3(1-b_{\bar{S}}^{\prime})H^{2}
\frac{v^{4}}{|S|^{2}}
+3(1-b_{N}^{\prime})H^{2}\frac{h^{2}}{4\lambda^{2}}\frac{|\phi|^{4}}{|S|^{2}}
\;.
\label{s1-9}
\end{eqnarray}
The above procedure can be justified as long as the $B-L$ breaking scale $v$
is larger than the Hubble parameter during inflation, because the
curvature around the minimum in Eq.~(\ref{s1-8}) coming from the D-term
potential is also of the order of the scale $v$.  We assume that this is
the case in the following discussion, which is, however, very plausible
since the $B-L$ breaking scale $v$ and
the Hubble parameter of inflation $H_{I}$ are most likely $v\gsim 10^{13}\GEV$
and $H_{I}\lsim10^{13}\GEV$.
As we will see later, the third term,
\begin{equation}
\frac{\sqrt{3}}{2}H\left(c_{\phi}^{\prime}\frac{h^{2}}{\lambda}
\frac{\phi^{4}}{S}+\rm{h.c.}\right)\;,
\label{HA}
\end{equation}
plays a crucial role in the D-term stopping case.
We will call this term ``Hubble A-term'' in the following sections.  
%%%%%%%%%%%%%%%%%%%%%%%%%%%%%%%%%
%%%%%%%%%%%%%%%%%%%%%%%%%%%%%%%%%%
\section{Evolution of the $\phi$ field during inflation}
%%%%%%%%%%%%%%%%%%%%%%%%%%%%%%%%%%
%%%%%%%%%%%%%%%%%%%%%%%%%%%%%%%%%%
In this section, we discuss the evolution of the $\phi$ field during the
inflation. As we will see later, in the D-term stopping case, the $\phi$ field settles down at
the $B-L$ breaking scale $v$ during the inflation.
This allows us to use the Hubble A-term as a torque to rotate the $\phi$
field after the inflation ends.  It will become clear in Section
\ref{Sec-BS} that the torque coming from this Hubble A-term is the
crucial ingredient to enhance the baryon asymmetry compared with the
global case as well as the  F-term stopping case.

First, let us investigate the potential for the $\phi$ field during 
the inflation in the range $|\phi|\lsim v$.
In this range, the $S$ field in Eq.~(\ref{s1-8}) is  expanded into
the following form:
\begin{equation}
|S|^{2}\simeq v^{2}-\frac{1}{8}|\phi|^{2}+{\cal{O}}\left(
\frac{|\phi|^{4}}{v^{2}}\right),\qquad \mbox{for} \quad |\phi|\lsim v
\;.
\label{s-expand}
\end{equation}
Then, by substituting Eq.~(\ref{s-expand}) into Eq.~(\ref{s1-9}),
we get the effective potential for the $\phi$ field in the 
range $|\phi|\lsim v$ in the following form:
\begin{eqnarray}
V\simeq&& V_{SB}+\frac{1}{2}\mu^{2}|\phi|^{2}+\frac{\sqrt{3}}{2}
\frac{H}{M}\left(c_{\phi}^{\prime}\phi^{4}+{\rm{h.c.}}\right)
+\frac{|\phi|^{6}}{4M^{2}}\nonumber\\
\nonumber\\
&&-H^{2}|\phi|^{2}+\frac{3}{8}\left(b_{S}^{\prime}-b_{\bar{S}}^{\prime}\right)
H^{2}|\phi|^{2}+\ldots
\;, 
\label{eVbv}
\end{eqnarray}
where the ellipsis denotes higher order terms in $|\phi|^{2}/v^{2}$.
Here, $M\equiv \lambda v/h^{2}$ corresponds to the same symbol used in 
Ref.~\cite{MM,AFHY,FHY} and the mass of the lightest neutrino is
written  as  $m_{\nu 1}=\left<H_{u}\right>^{2}/M$.

{}From Eq.~(\ref{eVbv}), we see that the effective potential is
almost the same as in the global case with some changes of coefficients
in the potential.  However, there is an important difference in the
Hubble-induced SUSY breaking mass terms.  We should not neglect the
contributions coming from the $S$ field.  In fact, we need the following
condition in order that the $\phi$ field develops a large expectation
value during the inflation:
\begin{equation}
b_{S}^{\prime}-b_{\bar{S}}^{\prime}\lsim \frac{8}{3}\;.  \label{bcon-1}
\end{equation}
If this is the case, the $\phi$ field has a negative Hubble mass term at
least in the range $|\phi|\lsim v$. Otherwise, the $\phi$ field is
driven toward the origin during the inflation because of the positive Hubble
mass term, and the MY leptogenesis  does not take place.

In the following discussion, we assume that Eq.~(\ref{bcon-1}) is
satisfied and discuss the scale where the flat direction is lifted. If
the balance point between the negative Hubble mass term and the F-term
potential $|\phi|^{6}/(4M^{2})$ in Eq.~(\ref{eVbv}) is below the $B-L$
breaking scale $v$, the $\phi$ field is stopped at this balance point,
\begin{eqnarray}
 |\phi|\simeq \sqrt{MH_{I}}< v
  \;,
  \label{Fst}
\end{eqnarray}
which corresponds to the F-term stopping case.  On the other hand, if
the condition
\begin{equation}
\sqrt{MH_{I}}\gsim v
\;,
\label{nonstop}
\end{equation}
is satisfied, the $\phi$ field can develop its expectation value as large 
as $v$.

%As we will see later, 
%the resultant baryon asymmetry is not enhanced in such a case and the MY
%leptogenesis results in the same conclusions as in 
%the global case~\cite{AFHY,FHY}.

Now, let us explain what happens when the expectation value of the
$\phi$ field grows as large as the $B-L$ breaking scale $v$.  At this
scale ({\it{i.e.}} $|\phi|\simeq v$), the expansion of the $S$ field
given in Eq.~(\ref{s-expand}) becomes invalid and above this scale we
must use another expansion of the $S$ field as follows:
\begin{equation}
|S|^{2}\simeq 4\frac{v^{4}}{|\phi|^{2}}+\frac{h^{2}}{2\lambda^{2}}|\phi|^{2}
+
\left[
{\cal{O}}\left( \frac{v^{4}}{|\phi|^{4}} \right)
+
{\cal{O}}\left( \frac{h^{2}}{\lambda^{2}} \right)
\right]^2
|\phi|^2
\qquad\mbox{for}\quad |\phi|\gsim v
\;.
\label{s-expand2}
\end{equation}
One might wonder whether the above expansion is reliable, since
Eq.~(\ref{s1-8}) is based on Eq.~(\ref{s1-6}) that may not be applicable
for $|\phi|\gg v$. Here, we first derive the conditions to fix the
$\phi$ field at the scale $v$ during the inflation, assuming this
expansion is effectively applicable at least for $|\phi|\sim v$. We will
justify later the validity of the obtained conditions by numerical
calculations.  

By substituting the expansion Eq.~(\ref{s-expand2}) into
Eq.~(\ref{s1-9}), we obtain the following Hubble-induced mass
term:\footnote{ Here, we assume $h^{2}/\lambda^{2}\ll 1$. If this is not
the case, we must include the Hubble mass term coming from the coupling
of the right-handed Majorana neutrino to the inflaton, which is the last
term in Eq.~(\ref{s1-9}).  }
\begin{equation}
V\simeq\left(-1+\frac{3}{4}(1-b_{\bar{S}}^{\prime})\right)H^{2}|\phi|^{2}
+{\cal{O}}\left(\frac{v^{4}}{|\phi|^{2}}H^{2}\right)
\qquad \mbox{for} \quad |\phi|\gsim v
\;.
\label{eVav}
\end{equation} 
Then, the $\phi$ field gets a positive Hubble mass term for $|\phi|\gsim
v$ if the following condition is satisfied:
\begin{equation}
b_{\bar{S}}^{\prime}\lsim -\frac{1}{3}\;.
\label{bcon-2}
\end{equation}
Therefore, if this is the case, the $\phi$ field cannot develop its
expectation value above the scale $v$ and hence it is fixed at the $B-L$
breaking scale $v$ during the inflation.

To summarize, the $\phi$ field, and hence $S$ and $\bar{S}$ fields also,
are stopped at the $B-L$ breaking scale $v$ during the inflation, if the
following conditions are satisfied,\footnote{The conditions for the
region $h^{2}/\lambda^{2} > 1$ are obtained by repeating the same
procedure.}
\begin{eqnarray}
\qquad b_{S}^{\prime}-b_{\bar{S}}^{\prime}\lsim \frac{8}{3},
\qquad \sqrt{M H_{I}}\gsim v,\qquad 
b_{\bar{S}}^{\prime}\lsim -\frac{1}{3}
\;.
\label{s1-11}
\end{eqnarray}
These are the conditions for the D-term stopping case.

\begin{figure}[t]
 \centerline{ {\psfig{figure=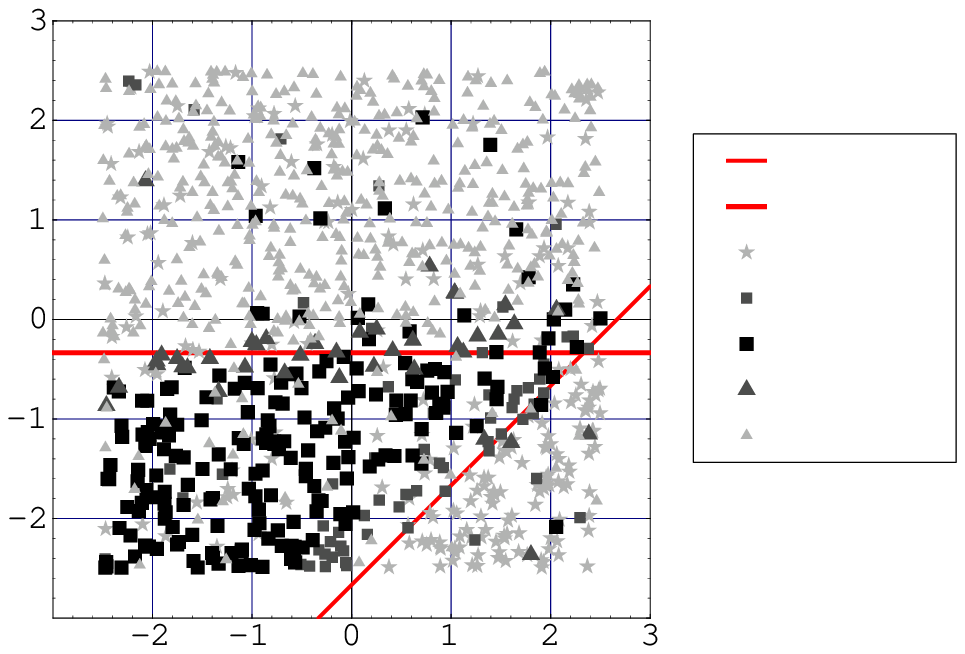,height=12cm}} }
\begin{picture}(20,0)
  \put(-30,190){$b_{\bar{S}}^{\prime}$}
  \put(150,1){$b_{S}^{\prime}$}
  \put(372,268){$b_{S}^{\prime}-b_{\bar{S}}^{\prime}=8/3$}
  \put(374,246){$b_{\bar{S}}^{\prime}=-1/3$}
  \put(375,223){$\phi/v<0.3$}
  \put(368,200){$0.3\leq\phi/v<1$}
  \put(368,178){$1\leq\phi/v<3$}
  \put(368,156){$3\leq\phi/v<10$} 
  \put(375,134){$10\leq\phi/v$}
\end{picture}
\\
\caption{The amplitudes of the $\phi$ field fixed during inflation which
are determined by numerical calculations. In this calculation, we have
used the full scalar potential in Eq.~(\ref{s1-5}) to follow the
evolution of the relevant fields.  Here, we have assumed that $
 H_{I}/v=0.1$, $3(1-b_{\phi}^{\prime})=-1$, $ h=10^{-4}$, $g,
\lambda,\eta=1$, and have randomly generated other coupling constants
$b_{Y},\;|c_{Y}|$ in the range $-2.5\leq b_{Y}\leq 2.5$ and $0\leq
|c_{Y}|\leq 2.5$, respectively.  Various symbols denote the $|\phi|$ to
$v$ ratio at the end of inflation.  We  see that the $\phi$ field
is, in fact, stopped at the $B-L$ breaking scale $v$ if the conditions
in Eq.~(\ref{s1-11}) are satisfied.  } \label{bcon}
\end{figure}

To check the conditions in Eq.~(\ref{s1-11}),
(especially that of the last condition $b_{\bar{S}}^{\prime}\lsim -
1/3$,) we have numerically solved the coupled equations of motions for
the relevant fields using the full scalar potential in Eq.~(\ref{s1-5}). 
We show the result in Fig.~\ref{bcon}, where the amplitude of the $\phi$
field at the end of the inflation is plotted in $b'_S$--$b'_{\bar{S}}$
plane. It is found that the conditions in Eq.~(\ref{s1-11}) well explain
the result of this numerical calculation.  In fact, the amplitude of the
$\phi$ field at the end of the inflation lies in the range $1\lsim
|\phi|/v\lsim 3$ in most of the parameter space where the conditions in
Eq.~(\ref{s1-11}) are satisfied.

%%%%%%%%%%%%%%%%%%%%%%%%%%%%%%%%%%
%%%%%%%%%%%%%%%%%%%%%%%%%%%%%%%%%%
%%%%%\section{Initial phase of the $\phi$ field}
%%%%%%%%%%%%%%%%%%%%%%%%%%%%%%%%%%
%%%%%%%%%%%%%%%%%%%%%%%%%%%%%%%%%%
As noted in the first paragraph of this section,
we can use the Hubble A-term potential [Eq.(\ref{HA})] as a torque to rotate the
$\phi$ field after the inflation ends in the D-term 
stopping case.  In the remaining part of this
section, we discuss this point in detail and compare it with the F-term
stopping case.

First, if the conditions in Eq.~(\ref{s1-11}) 
are satisfied ({\it{i.e.}}, the D-term stopping case), the curvature
of the potential for the $\phi$ field along the phase direction (which
is denoted by the symbol  $m^{2}_{\rm{phase}}$ )
is smaller than the Hubble parameter during the inflation.
This can be seen from the following relation:
\begin{equation}
m^{2}_{\rm{phase}}\simeq \frac{H_{I}}{M} |\phi|^{2}\simeq H_{I}^{2}
\left(\frac{|\phi|}{\sqrt{M H_{I}}}\right)^{2}
< H_{I}^{2}
\;.
\label{s1-12-1}
\end{equation}
Then, there is no reason to expect that the $\phi$ field sits down at
the bottom of the valley of the A-term potential in Eq.~(\ref{HA}) 
during the inflation.
Thus, unless there is an accidental fine tuning on the initial phase of
the $\phi$ field, it is generally displaced from the bottom of the
valley of this A-term potential when the inflation ends.  Therefore,
this A-term potential kicks the $\phi$ field along the phase direction
when the Hubble parameter becomes comparable with the curvature along
the phase direction $m^{2}_{\rm{phase}}$.  This is the reason why we can
use the Hubble A-term potential as a torque to rotate the $\phi$ field
and enhance the baryon asymmetry.

Next, we turn to the F-term stopping case.
In this case, the amplitude of the $\phi$ field is fixed at
$\sqrt{MH_{I}}$ during the inflation. [See Eq.~(\ref{Fst}).]
An important point is that
the curvature around the valley of the Hubble A-term potential,
$m^{2}_{\rm{phase}}$, is of the order of the Hubble parameter with the 
value of the $\phi$ field, $|\phi|\simeq \sqrt{MH_{I}}$:
\begin{equation}
m^{2}_{\rm{phase}}\simeq\frac{H_{I}}{M}|\phi^{2}|\simeq H_{I}^{2}
\;.
\label{s1-12}
\end{equation}
Therefore, the phase of the $\phi$ field settles down at the bottom of
the valley of this A-term potential during the inflation, and hence the
Hubble A-term cannot supply a torque to rotate the $\phi$ field.  In
this case, the relevant torque for the $\phi$ field only comes from the
ordinary A-term potential proportional to the gravitino mass.\footnote{
The initial phase of the $\phi$ field is generally displaced from the
bottom of the valley of this ordinary A-term potential unless the valley
of this A-term potential accidentally coincides with that of the Hubble
A-term potential.}  In this case, the MY leptogenesis with a gauged
$U(1)_{B-L}$ symmetry results in the same conclusions as in the global
case~\cite{AFHY,FHY}, in which the resultant baryon asymmetry is
proportional to the gravitino mass.  We stress that this leads to a
substantial suppression of the baryon asymmetry in gauge-mediation
models with a small gravitino mass.\footnote{ See the discussion in the
next section.}

Finally, we comment on the case where only the condition
$b_{\bar{S}}^{\prime}<-1/3$ in Eq.~(\ref{s1-11}) is not satisfied.  In
this case, the $\phi$ field has the negative Hubble-mass term even above
the $B-L$ breaking scale. We need the effective potential for the $\phi$
field above the scale $v$ in order to determine the scale at which the
$\phi$ field stops during the inflation.  However, there are so many
coupling constants and terms in Eq.~(\ref{s1-5}) which contribute to the
potential with comparable importance and hence it is very hard to get a
simple effective potential.  In spite of this complexity of the
potential, it is expected that the baryon asymmetry results in the range
between the following two cases.  If the curvature around the valley of
the Hubble A-term potential is as large as the Hubble parameter during
the inflation, $m^{2}_{\rm{phase}}\simeq H_{I}^{2}$, the baryon
asymmetry approaches what is obtained in the global
case~\cite{AFHY,FHY}.  On the other hand, if
$m^{2}_{\rm{phase}}<H_{I}^{2}$, the baryon asymmetry is enhanced and
probably the same as that in the D-term stopping case with which we are
mainly concerned in this paper.  In any case, we need detailed numerical
calculations using the full scalar potential in Eq.~(\ref{s1-5}) in
order to obtain the precise amount of the baryon asymmetry in this case,
which is beyond the scope of this work.

%%%%%%%%%%%%%%%%%%%%%%%%%%%%%%%%%%%%%%%%%%%%%%%%%%%%%%%%%%%%%%%%%%%%%%%%%%%%%%%%%%%%%%%%%%%%%%%%%%%%%%%%%%%%%%%%%%%%%%%%%%%%%%%%%%%%%%%%%%%%%%%%%%%%%%%%%%%%%%
\section{Baryon asymmetry}
%%%%%%%%%%%%%%%%%%%%%%%%%%%%%%%%%%%%%%%%%%%%%%%%%%%%%%%%%%%%%%%%%%%%%%%%%%%%%%%%%%%%%%%%%%%%%%%%%%%%%%%%%%%%%%%%%%%%%%%%%%%%%%%%%%%%%%%%%%%%%%%%%%%%%%%%%%%%%%
\label{Sec-BS} In this section, we calculate the baryon asymmetry.
First, let us discuss the D-term stopping case.  In this case, as
explained in the previous section, the $\phi$ field is fixed at the
$B-L$ breaking scale $v$ during the inflation, where the curvature along
the phase direction coming from the Hubble A-term
is smaller than the Hubble parameter $H_{I}$.  Thus, the
phase of the $\phi$ field is generally displaced from the bottom of the
valley of the Hubble A-term potential.

After the inflation ends, the universe becomes dominated by the
oscillating inflaton, and the scale factor of the expanding universe $R$
increases as $R\propto H^{-2/3}$ like in  the
matter-dominated universe~\cite{KolbTurner}. In the following
discussion, we assume that the production of the lepton number takes
place in this inflaton-oscillation dominated epoch, before the reheating
process of the inflation completes. (We will justify this assumption in
the next section.)

At first, the field value of the $\phi$ remains almost constant for a
while. This is because the amplitude of the $\phi$ field which gives the
potential minimum is mainly determined by the second line in
Eq.~(\ref{s1-9}) and it is given by $|\phi|\simeq v$ which is, of
course, independent of the Hubble parameter $H$.

Then, the Hubble A-term potential in Eq.~(\ref{HA}) 
kicks the $\phi$ field along the phase
direction when the Hubble parameter becomes comparable with the
curvature.  The Hubble parameter of this time is given by
\begin{eqnarray}
H_{ph}\simeq \frac{v^{2}}{M}
\quad&\Leftrightarrow&\quad
H_{ph}^{2}\simeq m^{2}_{\rm{phase}}\left(\simeq 
\frac{H_{ph}}{M}v^{2}\right)
\;.
\label{s2-13}
\end{eqnarray}
At this time ($H \simeq H_{ph}$), the phase of the $\phi$ field begins
to oscillate around the bottom of the valley of the Hubble A-term
potential. On the other hand, the amplitude of the $\phi$ field slowly
decreases after this time, according to
\begin{equation}
|\phi|\simeq \sqrt{M H}\;,
\label{s2-14}
\end{equation}
which is the balance point between the negative Hubble mass term and the
operator $|\phi|^{6}/(4 M^{2})$.

\begin{figure}[t]
 \centerline{ {\psfig{figure=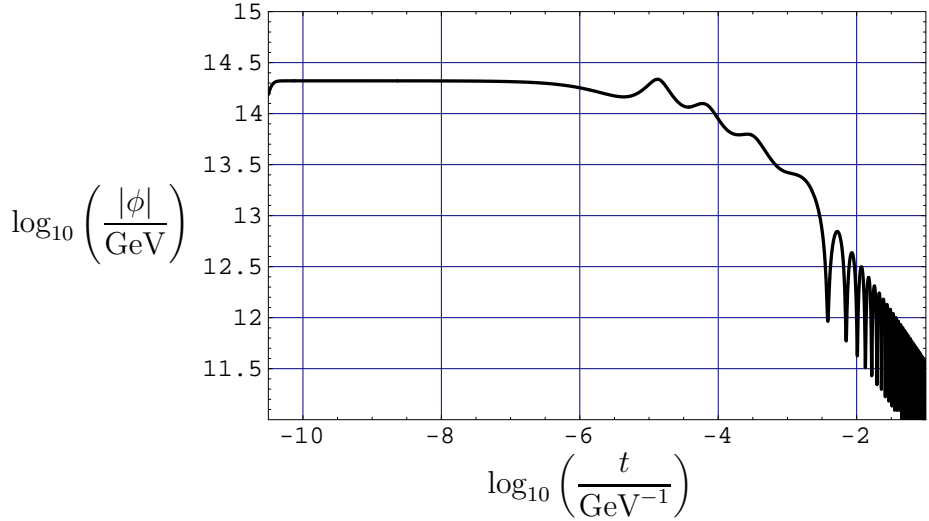,height=6cm}} }
\begin{picture}(1,0)
  \put(10,100){$\mbox{log}_{10}\left(\displaystyle{\frac{|\phi|}{\mbox{GeV}}}
\right)$}
  \put(190,1){$\mbox{log}_{10}\left(\displaystyle{\frac{t}{\mbox{GeV}^{-1}}}
\right)$}
 \end{picture}
\\
\caption{The evolution of the $\phi$ field estimated by a numerical
calculation. We assume that $M=3\times 10^{23}\GEV\;(\mbox{\it{i.e.}},\;
m_{\nu 1}=10^{-10}\EV),\;v=10^{14}\GEV,\;H_{I}=10^{12}\GEV,
\;m_{\phi}=10^{3}\GEV$. We see that the $\phi$ field
starts to move at $H=H_{ph} \simeq 3\times 10^4\GEV$ according
to the equation, $|\phi|\simeq\sqrt{MH}$. Here, $t=2/(3H)$ is the
cosmic time.  The $\phi$ field starts to oscillate around the origin at
$H_{osc}\simeq m_{\phi}$ since we neglect the thermal effects in this
calculation.}  \label{Amp}
\end{figure}

\begin{figure}[t]
 \centerline{ {\psfig{figure=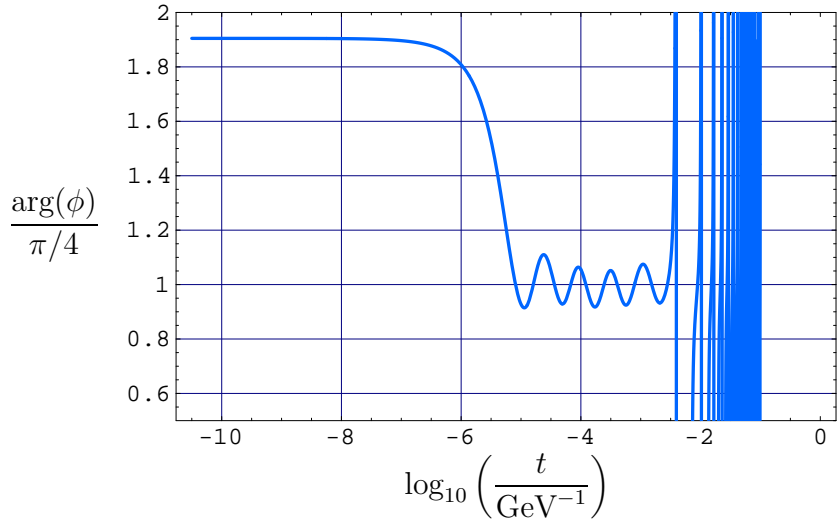,height=6cm}} }
\begin{picture}(0,0)
  \put(40,100){$\displaystyle{\frac{\mbox{arg}(\phi)}{\pi/4}}$}
  \put(190,1){$\mbox{log}_{10}\left(\displaystyle{\frac{t}{\mbox{GeV}^{-1}}}
\right)$}
 \end{picture}
\\
\caption{ The evolution of the phase of the $\phi$ field estimated by a
numerical calculation.  The parameters used in this figure are the same
as in Fig.~\ref{Amp}.  Here, we also defined that $c_{\phi}^{\prime}$ is
real, and hence the valleys coming from the Hubble A-term lie along
($\mbox{arg}(\phi)=\pi/4,\;3\pi/4,\;5 \pi/4,\;7\pi/4$).  We see that the
phase of the $\phi$ field starts to oscillate around the bottom of the
valley at $H\simeq H_{ph}$.  } \label{Phase}
\end{figure}

Since the $\phi$ field starts its oscillation along the phase direction
at $H=H_{ph}$, it has already had a large acceleration along the phase
direction before it starts coherent oscillation along the radius
direction around the origin when $H=H_{osc}$.  This $H_{osc}$ is
determined by thermal effects, soft SUSY breaking effects and $\mu$-term
in the same way as in Ref.~\cite{FHY}. For example, if the $\phi$ field
starts its oscillation due to the soft SUSY breaking mass in the true
vacuum $m_{\phi}$, it is given by $H_{osc} \simeq m_{\phi}$. The
evolution of the amplitude of the $\phi$ field and its phase obtained by
numerical calculations are shown in Fig.~\ref{Amp} and \ref{Phase}. In
these figures, $t$ is the cosmic time in the matter dominated universe
$t=2/(3H)$.  We see from these figures that the evolution of the $\phi$
field is well explained by above arguments.

We are now at the point to estimate the baryon asymmetry.  The lepton
number density is related to the $\phi$ field as follows:
\begin{equation}
n_{L}=\frac{1}{2}i\left(
\dot{\phi}^{*}\phi-\phi^{*}\dot{\phi}
\right)
\;,
\label{s2-15}
\end{equation}
where the overdot denotes a derivative with time.
The evolution of the $\phi$ field is described by the following equation
of motion:
\begin{equation}
\ddot{\phi}+3 H \dot{\phi}+\frac{\partial V}{\partial \phi^{*}}=0\;.
\label{s2-16}
\end{equation}
Then, we obtain, from Eq.~(\ref{s2-15}) and (\ref{s2-16}), the equation
of motion for the lepton number density as\footnote{Here, we neglect the
contribution coming from the ordinary A-term, since its contribution is
negligible.}
\begin{equation}
\dot{n}_{L}+3 H n_{L}=2\sqrt{3}\frac{H}{M}{\rm{Im}}\left(
c_{\phi}^{\prime}\phi^{4}\right)\;.
\label{s2-17}
\end{equation}
Here, we have used $|S|\simeq v$ for $H\lsim H_{ph}$.  This can be
easily integrated and we obtain the lepton number at time $t$ as
\begin{equation}
\left[R^{3}n_{L}\right](t)=\int^{t}dt R^{3}\;
2\sqrt{3}\frac{|c_{\phi}^{\prime}|H}{M}|\phi^{4}| \sin
\left(
{\rm{arg}}(c_{\phi}^{\prime})+4 {\rm{arg}}(\phi)
\right)
\;.
\label{s2-18}  
\end{equation}
In the regime $H_{osc}\lsim H\lsim H_{ph}$, the amplitude of the $\phi$
field decreases as $|\phi|\simeq \sqrt{MH}$ and the sign of ${\rm{sin}}$
in Eq. (\ref{s2-18}) changes with $1/H$ time scale. (See Fig.~\ref{Amp}
and \ref{Phase}.)  Therefore, from Eq. (\ref{s2-18}), one sees that the
total lepton number oscillates with $1/H$ time scale with almost
constant amplitude, since $R^3 H |\phi^4| \propto H \propto t^{-1}$ in
this regime.  As soon as the $\phi$ field starts its coherent
oscillation around the origin at $H=H_{osc}$, the total lepton number is
fixed, since the amplitude of the $\phi$ field decreases as fast as
$|\phi|\propto H$.  The lepton number density at time $t=2/(3H)$ is
written as
\begin{equation}
n_{L}(t)\simeq \frac{4}{\sqrt{3}}|c_{\phi}^{\prime}|M H^{2}\delta_{eff}\;,
\label{s2-19}
\end{equation}
where
$\delta_{eff}=\mbox{sin}\left(\mbox{arg}(c_{\phi}^{\prime})+4\mbox{arg}(\phi)
\right)$ is an ${\cal{O}}(1)$ effective CP-phase.  Note that the
Eq. (\ref{s2-19}) is correct as long as $H\lsim H_{ph}$, and hence the
resultant lepton-to-entropy ratio is totally independent of the time
when the $\phi$ field starts oscillations around the origin:\footnote{We
have also confirmed by numerical calculations that the final lepton
asymmetry $n_L / s$ is independent of the oscillation time $H_{osc}$.
See also the result for the F-term stopping case in Eq.~(\ref{s2-23}).}
\begin{equation}
\frac{n_{L}}{s}\simeq \frac{T_{R}M}{\sqrt{3}M_{*}^{2}}|c_{\phi}^{\prime}|
\delta_{eff}\;,
\label{s2-20}
\end{equation}
where $T_{R}$ is the reheating temperature of the inflation.  This
lepton asymmetry is partially converted into the baryon
asymmetry~\cite{LG-org} due to the ``sphaleron''
effects~\cite{sphaleron}, since it is produced before the electroweak
phase transition. The present baryon asymmetry is given
by~\cite{L-to-B}\footnote{In the present analysis, we neglect the
relative sign between the produced lepton and baryon asymmetries.}
\begin{equation}
\frac{n_{B}}{s}=\frac{8}{23}\frac{n_{L}}{s}\;.
\label{s2-21}
\end{equation}
Thus, after all, the present baryon asymmetry is given by 
\begin{eqnarray}
\frac{n_{B}}{s}&&\simeq \frac{8}{23\sqrt{3}}\frac{T_{R}M}{M_{*}^{2}}|c_{\phi}^{\prime}|\delta_{eff}\nonumber\\
\nonumber\\
&&\simeq 1.1\times 10^{-10}\left(\frac{T_{R}}{10^8\GEV}\right)
\left(\frac{10^{-6}\EV}{m_{\nu 1}}\right)|c_{\phi}^{\prime}|\delta_{eff}
\;.
\label{s2-22}
\end{eqnarray}
Here, we have used the relation $M=\left<H_{u}\right>^{2}/m_{\nu 1}$.

As stressed before, one sees that the resultant baryon asymmetry is
independent of the gravitino mass $m_{3/2}$ and the starting time of the
oscillation of the $\phi$ field, $H_{osc}$.  If the mass of the
gravitino is large enough $m_{3/2}\simeq {\cal{O}}(1\TEV)$, we can avoid
the cosmological gravitino problem even if the reheating temperature is
rather high $T_{R}\simeq 10^{10}\GEV$~\cite{G-prob-GRMSB}.  In such a
case, we see from the above equation that the mass of the lightest
neutrino $m_{\nu 1}\simeq 10^{-4}\EV$ is small enough to explain the
present baryon asymmetry.  On the other hand, if the mass of the
lightest neutrino is as small as $m_{\nu 1}\simeq 10^{-10}\EV$, we can
generate the required baryon asymmetry for low reheating temperatures as
$T_{R}\simeq 10^{4}\GEV$.  In such a low reheating temperature, we are
free from the overproduction of gravitinos even in gauge-mediated SUSY
breaking scenarios with the small gravitino mass
$m_{3/2}={\cal{O}}(100\kEV)$~\cite{g-prob-GMSB}.

\begin{figure}[t]
 \centerline{ {\psfig{figure=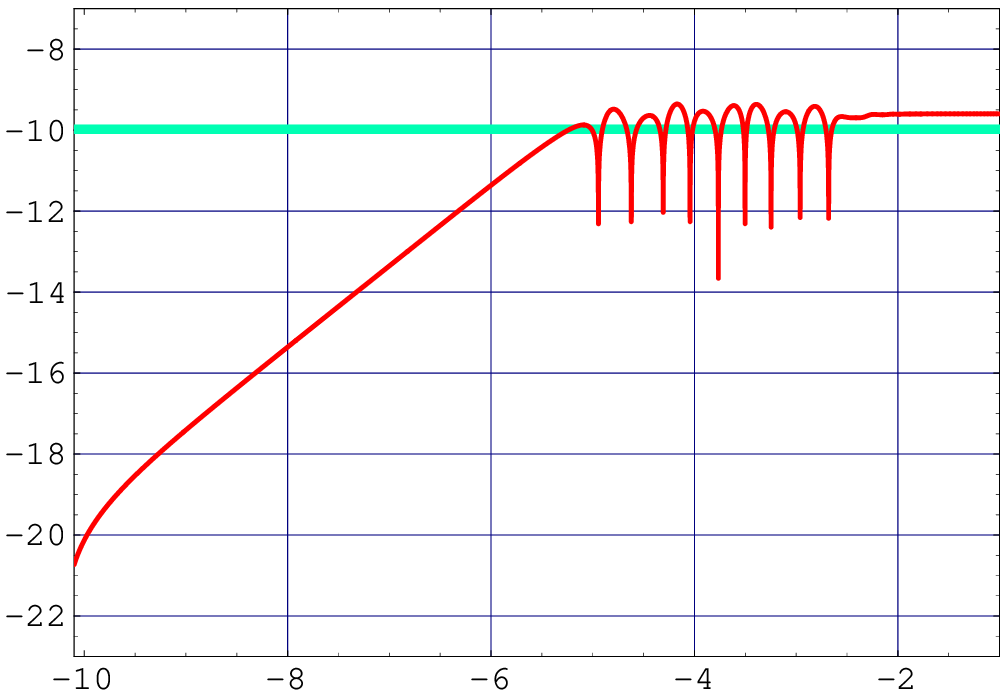,height=8cm}} }
\begin{picture}(20,0)
  \put(-5,130){$\mbox{log}_{10}\left(\displaystyle{\frac{n_{B}}{s}}\right)$}
  \put(190,1){$\mbox{log}_{10}\left(\displaystyle{\frac{t}{\mbox{GeV}^{-1}}}
\right)$}
 \end{picture}
\\
\caption{The evolution of the baryon asymmetry
$\displaystyle{\frac{n_{B}}{s}}=\frac{8}{23}\frac{n_{L}}{s}$ estimated
by a numerical calculation. Parameters used in this figure are the same
as in Fig.~\ref{Amp} and \ref{Phase}. Here, we also take
$T_{R}=10^{4}\GEV$.  The thick  line denotes the estimated value using
Eq. (\ref{s2-22}).  We can see that the asymmetry begins to oscillate
with constant amplitude at $H\simeq H_{ph}$ and it is fixed when the
$\phi$ field starts the oscillations around the origin. The analytical
estimation of the baryon asymmetry agrees well with the obtained value
by the numerical calculation.}  \label{Baryon}
\end{figure}

We show the evolution of the baryon asymmetry
$\displaystyle{\frac{n_{B}}{s}= \frac{8}{23}\frac{n_{L}}{s}}$ obtained
by a numerical calculation in Fig.~\ref{Baryon} and compare it with the
estimated value in Eq. (\ref{s2-22}).  {}From this figure we see that
the asymmetry begins to oscillate with almost constant amplitude at
$H=H_{ph}$ and it is fixed at the time when the $\phi$ field starts to
oscillate around the origin.  We confirm from this figure that the
resultant asymmetry is well explained by the arguments described in this
section.

Next, we briefly discuss the F-term stopping case, in which both the
evolution of the $\phi$ field and the resultant baryon asymmetry are
completely the same as those in the global case~\cite{FHY}.  In this
case, the $\phi$ field starts to move as soon as the inflation ends,
according to $|\phi|\simeq \sqrt{MH}$.  As mentioned in the previous
section, the Hubble-induced A-term cannot play a role to kick the phase
of $\phi$, since the flat direction field $\phi$ is trapped in the
valley of this Hubble-induced A-term in this case. The only available
A-term that can cause the motion of $\phi$ along the phase direction is
the ordinary A-term, which is proportional to the gravitino mass
$m_{3/2}$:
\begin{equation}
V_{SB}\ni \frac{m_{3/2}}{8M}\left(a_{m}\phi^4+{\rm{h.c.}}\right)
\;,
\label{globalA}
\end{equation}
where $a_{m}$ is an ${\cal{O}}(1)$ constant.
Thus, the equation of motion for the 
lepton number density in Eq.~(\ref{s2-17}) is
replaced by 
\begin{equation}
\dot{n}_{L}+3 H n_{L}=\frac{m_{3/2}}{2M}{\rm{Im}}\left(a_m \phi^{4}\right)\;.
\end{equation} 
By repeating the same procedure as in the D-term stopping case,
we obtain the following baryon asymmetry: 
\begin{eqnarray}
 \left.
  \frac{n_B}{s}
  \right|_{\rm F-term}
  \simeq
  \frac{2}{69}
  \frac{T_R M}{M_*^2}
  \left(
   \frac{m_{3/2}}{H_{osc}}
   \right)
  |a_{m}|
  \delta_{eff}
  \;.
\label{s2-23}
\end{eqnarray}
The factor
$(m_{3/2}/H_{osc})$ gives rise to a strong suppression of the resultant
baryon asymmetry for high reheating temperatures, since $H_{osc}$
becomes much larger than the soft mass $m_{\phi}$ in that
region~\cite{AFHY,FHY}.\footnote{The ``reheating temperature
independence of the baryon asymmetry'' in the global case comes from the
fact that $H_{osc}$ in Eq.(\ref{s2-23}) is proportional to $T_R$ or
$T_R^{3/2}$ in a large part of the parameter space, which leads to
$n_B/s \propto T_R^0$ or $T_R^{1/3}$~\cite{FHY}.} We also stress here
that this factor shows the strong suppression of the baryon asymmetry in
the gauge-mediated SUSY breaking scenario, since $m_{3/2} \ll m_{\phi} <
H_{osc}$.

%%%%%%%%%%%%%%%%%%%%%%%%%%%%%%%%%%%%%%%%%%%%%%%%%%%%%%%%%%%%%%%%%%%%%%%%%%%%%%%%%%%%%%%%%%%%%%%%%%%%%%%%%%%%%%%%%%%%%%%%%%%%%%%%%%%%%%%%%%%%%%%%%%%%%%%%%%%
\section{Other constraints}
\label{Sec-5}
%%%%%%%%%%%%%%%%%%%%%%%%%%%%%%%%%%%%%%%%%%%%%%%%%%%%%%%%%%%%%%%%%%%%%%%%%%%%%%%%%%%%%%%%%%%%%%%%%%%%%%%%%%%%%%%%%%%%%%%%%%%%%%%%%%%%%%%%%%%%%%%%%%%%%%%%%%%%%%
In the previous section, we found that the resultant baryon asymmetry is
enhanced in the D-term stopping case compared with the F-term stopping
case and also the global case.  However, in the D-term stopping case,
there are some conditions we must check, other than those in
Eq.~(\ref{s1-11}).  These are the conditions for avoiding the early
oscillation of the $\phi$ field.  Although the baryon asymmetry in the
D-term stopping case is independent of the $H_{osc}$ as long as $H_{osc}
< H_{ph}$, if the $\phi$ field starts its oscillation around the origin
before $H=H_{ph}$, {\it{i.e.}} $H_{osc}>H_{ph}$, the resultant baryon
asymmetry is strongly suppressed.\footnote{ In this case, the resultant
baryon asymmetry is inversely proportional to $H_{osc}^{2}$.}  The soft
SUSY breaking mass term of the $\phi$ field and the $\mu$-term cause the
early oscillation of the $\phi$ field when they satisfy\footnote{Note
that, in gauge-mediated SUSY breaking scenarios, the condition for
$m_{\phi}$ is absent if the messenger scale is lower than the breaking
scale of the $U(1)_{B-L}$. The existence of the $\mu$-term depends on
the energy scale of the dynamics which produces the $\mu$-term.}
\begin{equation}
m_{\phi},\;\; \mu > H_{ph}\simeq\frac{v^{2}}{M}\;.
\label{s3-23}
\end{equation}

Furthermore, thermal effects may cause the early oscillations.  A field
which couples to the $\phi$ field gets an effective mass $f_{k}|\phi|$,
where $f_{k}$ is a coupling constant to the $\phi$ field.  If the cosmic
temperature $T$ is larger than this effective mass, thermal fluctuations
of that field produce the thermal mass term for the $\phi$ field,
$c_{k}f_{k}^{2}T^{2}|\phi|^{2}$.  The list of $f_{k},\;c_{k}$ for the
$LH_{u}$ flat direction is given in Ref.~\cite{AFHY}.  Then, if one of
these thermal mass terms exceeds the Hubble parameter in the regime
$H>H_{ph}$, it causes the early oscillations of the $\phi$ field, and
then suppresses the baryon asymmetry substantially.  Namely, the
following condition is needed to avoid the early oscillation of the
$\phi$ field by the thermal mass terms:
\begin{equation}
H^2> \sum_{f_{k}v<T}c_{k}f_{k}^2T^2 \quad \mbox{for}\;\;H>H_{ph}\;.
\label{s3-24}
\end{equation}

\begin{figure}[t]
 \centerline{ {\psfig{figure=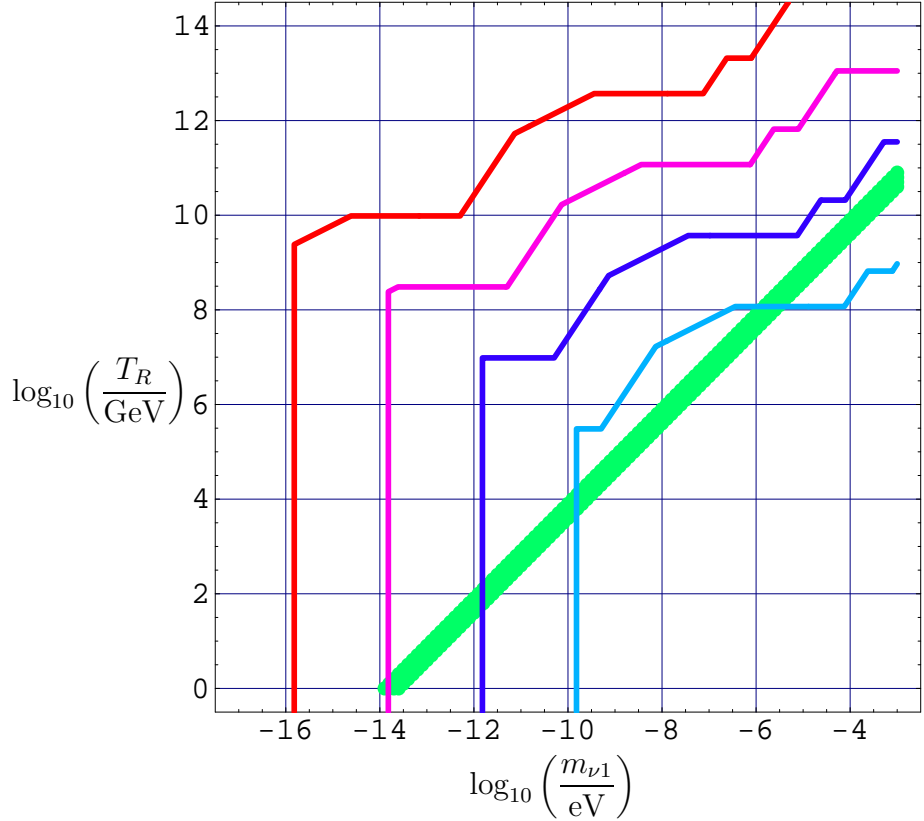,height=10cm}} }
\begin{picture}(0,0)
  \put(17,150){$\mbox{log}_{10}\left(\displaystyle{\frac{T_{R}}{\mbox{GeV}}}
\right)$}
  \put(190,1){$\mbox{log}_{10}\left(\displaystyle{\frac{m_{\nu 1}}{\mbox{eV}}}
\right)$}
 \end{picture}
\\
 \caption{The plot of the parameter regions which are free from the
 early oscillation in $m_{\nu 1}-T_{R}$ plane.  The regions below the
 four solid lines are free from the early oscillation. These four lines
 correspond to the breaking scale of the $U(1)_{B-L}$,
 $v=10^{16},\;10^{15},\;10^{14},\;10^{13}\; (\mbox{GeV})$ from left to
 right, respectively. The shaded region corresponds to the present
 baryon asymmetry, $n_{B}/s\simeq (0.4-1)\times 10^{-10}$. We have taken
 $\delta_{eff},\;|c_{\phi}^{\prime}|=1$ in this figure. Here, we assume
 the existence of $m_{\phi}$ or the $\mu$-term when the amplitude of the
 $\phi$ field is of the order of the $U(1)_{B-L}$ breaking scale
 $v$. {}From this figure, we see that if we take the breaking scale of
 the $U(1)_{B-L}$, $v\gsim 10^{14}\GEV$, the early oscillation of the
 $\phi$ field $H_{osc}>H_{ph}$ can be avoided in most of the regions of
 parameter space.}
 \label{Fig-Early}
\end{figure}

There is another thermal effect which was pointed out in
Ref.~\cite{AnisimovDine}.  Along the $LH_{u}$ flat direction, $SU(3)$
gauge symmetry remains unbroken, and hence, gluons and gluinos are
massless.  Furthermore, the down type (s)quarks also remain massless
since they have no coupling to the $\phi$ field.  These light fields
produce the free energy which depends on the $SU(3)$ gauge coupling
constant. We obtain the effective potential $V\propto g_{S}^{2}T^{4}$ at
two loop level~\cite{T4term}, where $g_{S}$ is the gauge coupling
constant of the $SU(3)$.  At first sight, there seems no dependence on
the $\phi$ field in this free energy. However, the up type (s)quarks get
large masses from the couplings to the $\phi$ field $y_{u}$, and if
$y_{u}|\phi|>T$, they decouple and change the trajectory of the running
coupling constant of the $SU(3)$.  This effect produces the effective
potential for the $\phi$ field,\footnote{ For non-abelian gauge
symmetries, decouplings of matter particles give always positive
contributions, and the net contribution from decouplings of gauge bosons
and gauginos is always negative.  The $LH_{u}$ flat direction has only
positive contributions for $SU(3)$ gauge symmetry.  On the other hand,
other flat directions such as $\bar{u}\bar{d}\bar{d},\;LL\bar{e},\ldots$
receive large negative contributions and probably have negative
thermal-log term.  }
\begin{equation}
\delta V\simeq a_{g}\alpha_{S}^{2}T^{4}{\rm{log}}\left(\frac{|\phi|^{2}}
{T^{2}}\right)\;,
\label{s3-25}
\end{equation}
where $a_{g}$ is a constant a bit larger than unity and
$\alpha_{S}\equiv g_{S}^{2}/4\pi$~\cite{FHY,AnisimovDine}.
Then, the following condition is required
to avoid the early oscillation caused by this potential:
\begin{equation}
H^{2}>\frac{a_{g}\alpha_{S}^{2}T^4}{v^{2}}\quad \mbox{for}\;\;H>H_{ph}
\;.
\label{s3-26}
\end{equation}
{}From Eq.(\ref{s3-24}) and (\ref{s3-26}),
we obtain the following condition to avoid the early oscillation by
thermal effects:
\begin{equation}
T_{R}<\mbox{min}\left[
\mbox{min}_{k}\left\{
\mbox{max}\left(
\frac{f_{k}v^{3/2}}{c_{k}^{1/4}M_{*}^{1/2}},\;\;\frac{v^3}{c_{k}f_{k}^2
(M_{*}M^3)^{1/2}}
\right)\right\}
,\;\;\frac{v^2}{a_{g}^{1/2}\alpha_{s} (M_{*} M)^{1/2}}
\right]\;,
\end{equation}
where, we have used the fact that the cosmic temperature
behaves as $T=(HT_{R}^{2}M_{*})^{1/4}$ before the reheating process of
the inflation ends\cite{KolbTurner}.  Taking all of these effects into
account, we get the allowed regions which are free from the early
oscillations as in Fig.~\ref{Fig-Early}.  The regions below four solid
lines are free from the early oscillations.  These four lines correspond
to the breaking scale of the $U(1)_{B-L}$,
$v=10^{16},\;10^{15},\;10^{14},\;10^{13}\GEV$ from left to right,
respectively.  The vertical parts of these lines come from the condition
in Eq. (\ref{s3-23}).  Here, we assume the existence of $m_{\phi}$ or
$\mu$-term when the amplitude of the $\phi$ field is of the order of the
$B-L$ breaking scale.  The shaded region denotes the present baryon
asymmetry, $n_{B}/s\simeq (0.4-1)\times 10^{-10}$.  {}From the
Fig.~\ref{Fig-Early}, we see that the early oscillations can be easily
avoided, especially when the $B-L$ breaking scale satisfies $v\gsim
10^{14}\GEV$.\footnote{In the regions above the four solid lines,
the estimation of $H_{osc}$ is different from that in Ref.~\cite{FHY}.}

Finally, we here check the assumption that the production of the lepton
asymmetry takes place during the inflaton-oscillation dominated era,
{\it{i.e.}}, before the reheating process of the inflation completes.
Then, we have the following constraint;
\begin{eqnarray}
 H_{osc} > \Gamma_I \simeq 
  \left(
   \frac{\pi^2 g_*}{90}
   \right)^{1/2}
   \frac{T_R^2}{M_*}
   \;,
   \label{s3-27}
\end{eqnarray}
where $\Gamma_I$ is the decay rate of the inflaton $I$ and $g_*$ denotes
the number of relativistic degrees of freedom, which is $g_*(T)\simeq
200$ in the MSSM for $T \gg 1\TEV$. As long as there exists a soft mass
$m_{\phi}$ or the $\mu$-term at high energy scale, $H_{osc}$ is at least
larger than that. Thus, the above constraint is indeed satisfied for
$T_R \lsim 10^{10}\GEV\times (m_{\phi}\,{\rm or}\,\mu / 1
\TEV)^{1/2}$. Furthermore, the Hubble parameter of the oscillation time
$H_{osc}$ becomes much larger for higher reheating temperatures, because
of the thermal effects. For example, the $H_{osc}$ always satisfies the
following relation~\cite{FHY};
\begin{eqnarray}
 H_{osc} \ge \alpha_S T_R
  \left(
   a_g
   \frac{M_*}{M}
   \right)^{1/2}
   \;.
\end{eqnarray}
Thus, the constraint given in Eq.~(\ref{s3-27}) is satisfied as long as
\begin{eqnarray}
 T_R \lsim 10^{15}\GEV
  \times
  \left(
   \frac{m_{\nu 1}}{10^{-8}\EV}
   \right)^{1/2}
   \;,
\end{eqnarray}
which is the case in all the relevant parameter space in the present
analysis. (See Fig.~\ref{Fig-Early}.) Therefore, the assumption
$\Gamma_I < H_{osc}$  is justified.

%%%%%%%%%%%%%%%%%%%%%%%%%%%%%%%%%%%%%%%%%%%%%%%%%%%%%%%%%%%%%
%%%%%%%%%%%%%%%%%%%%%%%%%%%%%%%%%%%%%%%%%%%%%%%%%%%%%%%%%%%%%
\section{Q-ball problem}
%%%%%%%%%%%%%%%%%%%%%%%%%%%%%%%%%%%%%%%%%%%%%%%%%%%%%%%%%%%%%%
%%%%%%%%%%%%%%%%%%%%%%%%%%%%%%%%%%%%%%%%%%%%%%%%%%%%%%%%%%%%%
One of the physical reasons to make the leptogenesis via $LH_{u}$ flat
direction so special among other baryo/leptogenesis
scenarios using the AD mechanism is that it is free from the
Q(L)-ball~\cite{Coleman:1985ki} problem.  The coherent oscillation of
the $\phi$ field is unstable with spatial perturbations if the potential
of the $\phi$ field is flatter than the quadratic potential.\footnote{
In this section, we will use the same symbol $\phi$ to
denote the field which parameterizes a general flat direction.  }  If
this is the case, the coherent oscillation of the $\phi$ field fragments
into non-topological solitons, Q-balls~\cite{Kusenko:1998si}.  By the
recent detailed analysis by lattice simulations, it becomes clear that
almost all of the charges carried by the $\phi$ field are absorbed into
these Q-balls~\cite{Qsimulation}, and the present baryon asymmetry must
be provided by the decay of the Q-balls, not by the direct decay of the
$\phi$ field.  The formation of Q-balls is a generic feature of the AD
baryo/leptogenesis regardless of the mediation mechanism of SUSY
breaking.

In gravity-mediated SUSY breaking scenarios, the potential of the $\phi$
field is slightly flatter than the quadratic potential due to the
running of the soft mass of the $\phi$ field coming from gaugino
loops~\cite{Enqvist,INSTA}.  In this case, the Q-ball is generally
unstable.  A difficulty arises from its long life time.  If the decay
temperature of the Q-balls is well below the freeze out temperature of
the lightest supersymmetric particle (LSP), the amount of the LSP cold
dark matter can be written as
\begin{equation}
\Omega_{\chi}=3\left(\frac{N_{\chi}}{3}\right)
f_{B}\left(\frac{m_{\chi}}{m_{n}}\right)\Omega_{B}
\;,
\label{s4-27}
\end{equation}
where $N_{\chi}$ is the number of LSP's produced per baryon number,
which is at least 3, and $f_{B}\simeq 1$ is the fraction of baryon
number stored in the form of Q-balls, and $m_{n}$ and $m_{\chi}$ are the
nucleon mass and the neutralino $(\chi)$ LSP mass, respectively.
$\Omega_{X}$ denotes the ratio of the energy density of $X$ to the
critical density of the present universe.  {}From this relation, it is
clear that the late time Q-ball decay leads to the over production of
the LSP cold dark matter.  In fact, this late time decay of Q-balls
invalidates most of the AD baryo/leptogenesis scenarios in
gravity-mediated SUSY breaking models.\footnote{The present status of AD
baryo/leptogenesis in gravity-mediated SUSY breaking models is
summarized in Ref.~\cite{ADwB-L}.  There, we also proposed an
interesting AD baryo/leptogenesis model with a gauged $U(1)_{B-L}$
symmetry, which is an almost unique solution to solve the Q-ball problem
in gravity-mediation models except the present leptogenesis via $LH_{u}$
flat direction. }

On the other hand, in gauge-mediated SUSY breaking scenarios, the soft
SUSY breaking mass of the $\phi$ field vanishes above the messenger
scale, and the $\phi$ field has only logarithmic potential.  This fact
leads to the formation of Q-balls.  The Q-balls in gauge-mediated SUSY
breaking scenarios have very different characters compared with those in
gravity-mediated SUSY breaking models.  The effective mass of the Q-ball
per baryon number is proportional to $Q^{-1/4}$~\cite{Dvali:1998qv},
where $Q$ is the charge of the Q-ball.  Therefore, if the charge of the
produced Q-ball is large, the Q-balls become completely stable and they
remain as a cold dark matter in the universe~\cite{Kusenko:1998si}.
Once we fix the flat direction for the AD baryo/leptogenesis, we can
estimate the size of the produced Q-ball and its number density by using
the result of the recent numerical calculations in
Ref.~\cite{Q-prob-GMSB}.  We find that the produced Q-balls overclose
the universe in almost all of the regions of parameter space.  Though
there remains some tiny regions in parameter space in which the produced
Q-balls do not overclose the universe, most part of such regions are
already experimentally excluded~\cite{Q-prob-GMSB}. If we use the
leptonic flat directions, such as $LL\bar{e}$, the produced Q(L)-balls
can decay into neutrinos, and hence they do not overclose the
universe. Unfortunately, however, we can only use the leptonic charges
which evaporates before the electroweak phase transition, because we
must convert the lepton asymmetry to the baryon asymmetry by ``sphaleron
effects''~\cite{sphaleron}.  As a result, we find that the present
baryon asymmetry can be explained only in extremely small regions of the
parameter space.\footnote{The authors thank S.~Kasuya and M.~Kawasaki
for useful discussion.}

How about the $LH_{u}$ flat direction?  The soft SUSY breaking mass of
the $\phi$ field has no contribution from gluino loops, but, on the
other hand, it has a big opposite contribution from large top Yukawa
coupling, which makes the potential for the $\phi$ field steeper than
quadratic potential.  Therefore, the gravity-mediation type Q-balls are
not formed in the $LH_{u}$ flat direction~\cite{INSTA}.  Furthermore,
the $LH_{u}$ flat direction is the unique flat direction available for
the AD baryo/leptogenesis which has a SUSY mass term, $\mu$-term.  Thus,
once we assume the existence of the $\mu$-term even in the high energy
scale relevant for the MY leptogenesis, there always exists the
quadratic potential for the $\phi$ field and the formation of Q-balls is
avoided even in the gauge-mediated SUSY breaking scenarios. (This is
possible if the $\mu$-term is generated by an expectation value of a
nonrenormalizable operator.)

Therefore, the MY  leptogenesis is free from the
Q-ball problem not only in gravity-mediated SUSY breaking scenarios but
also in gauge-mediated SUSY breaking scenarios.\footnote{If the $\phi$
field starts to oscillate around the origin due to the thermal-log term
$\delta V=a_{g}\alpha_{s}^{2}T^{4}
\mbox{log}\left(\displaystyle{|\phi|^{2}/T^{2}}\right)$, gauge-mediation
type Q-balls are formed. Fortunately, however, in such a case, the
Q-balls produced by the thermal-log term are small enough to evaporate
well above the electroweak phase transition. Therefore, the thermal-log
term does not harm the leptogenesis via $LH_{u}$ flat direction.}
%%%%%%%%%%%%%%%%%%%%%%%%%%%%%%%%%%%%%%%%%%%%%%%%%%%%%%%
%%%%%%%%%%%%%%%%%%%%%%%%%%%%%%%%%%%%
\section{Discussion and conclusions}
%%%%%%%%%%%%%%%%%%%%%%%%%%%%%%%%%%%%
%%%%%%%%%%%%%%%%%%%%%%%%%%%%%%%%%%%%%%%%%%%%%%%%%%%%%%%
In this paper, we have investigated the supersymmetric leptogenesis via
$LH_{u}$ flat direction originally suggested by Murayama and Yanagida
(MY)~\cite{MY} in the presence of a gauged $U(1)_{B-L}$ symmetry.
We have shown that if the flat direction field $\phi$ is stopped by the 
F-term potential, the situation is almost the same as that in the 
global case where the $U(1)_{B-L}$ is not gauged~\cite{FHY}.

Interestingly, however, we have found that the resultant baryon
asymmetry in the D-term stopping case is enhanced compared with the
global case.  In this case, the $\phi$ field is stopped at the $B-L$ breaking scale,
where the curvature along the phase direction coming from the
Hubble-induced A-term is smaller than the Hubble parameter.  This allows
us to use a much bigger torque to rotate the $\phi$ field, since the
$\phi$ field is initially displaced from the bottom of the valley of the
Hubble A-term potential unless we tune the initial phase of the $\phi$
field.  We have shown that the baryon asymmetry is proportional to the
reheating temperature of the inflation, but it is completely independent
of the size of the gravitino mass.  As a result, if high reheating
temperatures $T_{R}\sim 10^{10}\GEV$ are available, the mass of the
lightest neutrino, $m_{\nu1}\sim 10^{-4}\GEV$, is sufficiently small to
explain the present baryon asymmetry.  Furthermore, the gravitino mass
independence of the resultant baryon asymmetry gives us a great
advantage in gauge-mediated SUSY breaking scenarios.  If the mass of the
lightest neutrino is as small as $m_{\nu 1}\simeq 10^{-10}\EV$, we can
explain the present baryon asymmetry even in the case where the
gravitino mass is ${\cal{O}}(100\mbox{keV})$, which may be directly
tested in future experiments~\cite{MakiOrito}.  We stress that our
scenario is the first minimal framework which can explain the present
baryon asymmetry in gauge-mediated SUSY breaking scenarios with such a
small gravitino mass.

In addition, the MY leptogenesis via $LH_{u}$ flat direction 
is very special in the sense that it is completely 
free from the Q-ball problem.
In gauge-mediation scenarios, this requires the existence of the
$\mu$-term (SUSY-invariant mass for Higgs multiplets) in the high energy scale relevant for the  leptogenesis,
which may imply the high energy origin of the $\mu$-term in
gauge-mediated SUSY breaking scenarios.
The generation of the correct size of the $\mu$-term has been one 
of the most difficult problems in gauge-mediated SUSY breaking models.
The presence of the  baryon asymmetry in our universe can be regarded
as an interesting implication for the origin of the $\mu$-term.

Finally, we briefly comment on the prediction on the rate of the
$0\nu\beta\beta$ decay.  As in the global case, the present scenario
also requires a hierarchical neutrino mass spectrum to generate the
present baryon asymmetry even in the D-term stopping case. This results
in almost the same prediction on the rate of the $0\nu\beta\beta$ decay
as that in the global case~\cite{FHY}.  This consistency check of the
present leptogenesis scenario will be given in future $0\nu\beta\beta$
decay experiments~\cite{GENIUS}.
%%%%%%%%%%%%%%%%%%%%%%%%%%%%%%%%%%%%%%%%%%%%%%%%%%%%%%%%%%%%%%%%%%%
\section*{Acknowledgments}
%%%%%%%%%%%%%%%%%%%%%%%%%%%%%%%%%%%%%%%%%%%%%%%%%%%%%%%%%%%%%%%%%%%
The authors are grateful to S.~Kasuya and M.~Kawasaki for useful
discussion.  M.F. and K.H. thank the Japan Society for the Promotion of
Science for financial support.  This work was partially supported by
``Priority Area: Supersymmetry and Unified Theory of Elementary
Particles (\# 707)'' (T.Y.).
%\normalsize
%\appendix
%%%%%%%%%%%%%%%%%%%%%%%%%%%%%%%%%%%%%%%
%%%%%%%%%%%%%%%%%%%%%%%%%%%%%%%%%%%%%%%%
%\section{The evolution of $H_{d}$}
%\label{sec:appendix}

%%%%%%%%%%%%%%%%%%%%%%%%%%%%%%%%%%%%%%%
%%%%%%%%%%%%%%%%%%%%%%%%%%%%%%%%%%%%%%

%%%%%%%%%%%%%%%%%%%%%%%%%%%%%%%%%%%%%%%%%%%%%%%%%%%%%%%%%%%%%%%%%%%%
%
% \clearpage
%
%%%%%%%%%%%%%%%%%%%%%%%%%%%%%%%%%%%%%%%%%%%%%%%%%%%%%%%%%%%%%%%


\end{document}